\documentclass[conference,final]{IEEEtran}

\IEEEoverridecommandlockouts                              

\usepackage{graphicx} 
\usepackage{float}
\usepackage[utf8x]{inputenc}
\usepackage{eurosym}
\usepackage{fontawesome}
\usepackage{color}
\usepackage{booktabs}
\usepackage{tabularx}
\usepackage{nohyperref}  
\usepackage{url}  

\usepackage{import}

\newcommand{\footurl}[1]{\footnote{\url{#1}}}

\title{
Smart Street Lights and Mobile Citizen Apps for Resilient Communication in a Digital City 
}

\author{
\IEEEauthorblockN{
Lars Baumg\"{a}rtner\IEEEauthorrefmark{1},
Jonas H\"{o}chst\IEEEauthorrefmark{2},
Patrick Lampe\IEEEauthorrefmark{2},
Ragnar Mogk\IEEEauthorrefmark{1},\\
Artur Sterz\IEEEauthorrefmark{2},
Pascal Weisenburger\IEEEauthorrefmark{1},
Mira Mezini\IEEEauthorrefmark{1},
Bernd Freisleben\IEEEauthorrefmark{2}
}\\

\IEEEauthorblockA{
\IEEEauthorrefmark{1} Technische Universit\"{a}t Darmstadt, FB 20, D-64289 Darmstadt, Germany\\
E-mail: \{baumgaertner, mogk, weisenburger, mezini\}@cs.tu-darmstadt.de \\
\IEEEauthorrefmark{2}Philipps-Universit\"{a}t Marburg, FB 12, D-35032 Marburg, Germany\\
E-mail: \{hoechst, lampep, sterz, freisleb\}@informatik.uni-marburg.de\\
}
}

\begin{document}

\maketitle

\thispagestyle{empty}
\pagestyle{empty}

\begin{abstract}
While information and communication technology is crucial for the operation of urban infrastructures and the well-being of its inhabitants, current technology is quite vulnerable to disruptions of various kinds. In future smart cities, a more resilient urban infrastructure is imperative to handle the increasing number of hazardous situations.
We present a novel resilient communication approach based on smart street lights as part of the public infrastructure.  It supports people in their everyday life and adapts its functionality to the challenges of emergency situations. Our approach relies on various environmental sensors and in-situ processing for automatic situation assessment, and a range of communication mechanisms 
for maintaining a communication network. 
Furthermore, resilience is not only achieved based on infrastructure deployed by a digital city's municipality, but also based on integrating citizens through software that runs on their mobile devices. 
Web-based zero-installation and platform-agnostic apps can switch to device-to-device communication to continue benefiting people even during a disaster situation. Our approach, featuring a covert channel for professional responders and a zero-installation app, is evaluated through a prototypical implementation based on a commercially available street light.

\end{abstract}

\IEEEpeerreviewmaketitle


\section{Introduction}
\label{sec:intro}

In 2050, about two thirds of the world's population will live in cities -- after 30 percent in 1950 and 50 percent in 2010. 
This will affect an estimate of nearly 7 billion people.
For some years now, a transformation from cities that have grown over centuries to digital cities is in progress.
By a digital city we mean the complex networked cyber-physical system that extends the physical functions and activities in the city through a digital space created by modern information and communication technologies (ICT) and enables a multitude of services \cite{cocchia2014smart}. 
Significant system fluctuations, shock or crisis events (collectively referred to as crises) call this state into question by impairing the availability of (critical) infrastructures. Exemplary crisis scenarios are: a long-lasting power failure, a natural disaster like an earthquake, a massive cyber attack on communication networks,
and a terrorist attack or rampage.
Urban resilience\footnote{According to the definition of the United Nations Office for Disaster Risk Reduction. http://www.unisdr.org/we/inform/terminology} is the opposite pole and describes a city's ability to cope with crises and dangerous situations and to recover from the effects promptly and sustainably and to restore the necessary basic structures.

In this paper, we present a novel resilient communication approach 
for a digital city.\footnote{Source code is available at: 
\url{https://github.com/stg-tud/emergencity_demo}} It integrates citizens, their various devices, and public infrastructure into a whole. Smart street lights are used for sensing critical events and for forming a communication backbone. While smart street lights have many benefits, such as weather and air quality measurement plus public Internet access and tourist information during non-crisis times, they can functionally morph into something different in the event of an emergency. Here, additional sensors are activated and sensing can be performed with a higher sampling rate, plus live video feeds can be used. The sensors (especially a camera) can be used to automatically detect critical situations in-situ prior to an emergency alarm. Furthermore, a hidden covert channel is used for secure communication with a command center during crises. In addition to the public web services that can be accessed via the street light, we also provide a zero-installation app that works on Android and iOS that can be used to access all information in a convenient way. This app continues to work during a crisis and without the Internet by transferring information from device to device. Furthermore, the app enables users to share information and data (e.g., images) with other users. Providing such benefits for users not only during a crisis but also during everyday life is key to real-world adoption of such new technologies.
We make the following contributions:

\begin{itemize}
\item We present a novel smart street light for environmental sensing during everyday life and 
functional morphing during crisis events.
\item We present a novel covert channel for communication during crises.
\item We present a novel platform agnostic zero-installation mobile app that can handle connection loss and provide information and assistance in everyday life as well as during a crisis.
\end{itemize}

The paper is organized as follows.
Section \ref{sec:rw} discusses related work. 
In Section \ref{sec:design}, we present requirements and  designs decisions.
Section~\ref{sec:impl} discusses implementation issues.
Section \ref{sec:eval} presents experimental results.
Section \ref{sec:concl} concludes the paper and outlines areas of future work.


\section{Related Work}
\label{sec:rw}

Reliability has been a major research focus in distributed systems for computational cluster environments, such as MapReduce~\cite{Dean2008,Lammel20081}.
Modern languages in that area, such as Flink~\cite{Alexandrov2014TheSP,Carbone2015LAS}, provide high-level abstractions to programmers, which provide completely automatic fault-tolerance, synchronization, and data management.
Techniques such as GSP~\cite{Burckhardt2015} provide clear semantics for eventually consistent systems, without sacrificing developer convenience.
However, none of these solutions apply to emergency scenarios in a digital city, where failed hardware cannot be immediately replaced by idle compute instances. Instead, we have to deal with degraded information and unreliable communication.
Actor languages have been successfully applied to mobile ad-hoc systems~\cite{VanCutsem2014}, but do not provide support for automatic distribution of relevant information -- developers have to design their own solutions instead, a task that often is too costly for rare but important emergency cases.
To bridge the gap between automatic solutions and solutions applicable to emergency scenarios, we follow a line of research discussed by Hellerstein et al.~\cite{Hellerstein2019} who show that systems built on monotonic operations can provide consistency as well as availability even in the worst network conditions.

Another part of our work is to monitor data from sensors. In the literature, there are many examples for sensing platforms like the OpenSense project that provides an open platform for community-driven environmental monitoring and also makes the results publicly available for interested researchers and citizens \cite{aberer2010opensense}.
Other projects focus mainly on air quality measurements in urban environments.  Citi-Sense-MOB \cite{castell2015mobile} and AirSenseEUR\footnote{https://airsenseur.org/website/} \cite{gerboles2015airsenseur} are prominent examples. Their main drawback is the sole focus on a single task and their dependence on infrastructure for communication and power supply.
Often, Arduino or Raspberry Pi computers are the foundations for open sensor platforms ~\cite{llamas2017open}.
One such platform is the educational and open data senseBox project\footnote{https://www.sensebox.de} that can easily be used for citizen science projects \cite{wirwahn2015usability}. 
The open data and citizens science spirit is also present in the MCU-based air quality measurement project Luftdaten\footnote{https://www.luftdaten.info} \cite{froschle2017engineering}.
Sethi et al. \cite{sethi2018robust} present a robust solution for monitoring ecosystems. It is based on Raspberry Pi computers that measure sensor data over time to track changes in nature and various ecosystems. The focus of the work is to provide a cost efficient autonomous solution for monitoring ecosystems in the wild. The presented approach satisfies most of our requirements, but we had more size constraints to bring our solution inside a street light.
Aide et al. \cite{aide2013real} present a solution to acoustically monitor the environment with an iPod Touch (2G). This solution only measures events using one sensor, a microphone, and is not designed to fit our needs to monitor different type of sensors to get a situational overview during a crisis.

To summarize, several projects tackle at least part of our problem, but most solutions do not take existing urban infrastructure into account and consider equipping it with additional parts. In this case, we have size constraints for retrofitting, but we can benefit from the existing infrastructure as well. 
Most of the existing solutions focus on a main task to complete, e.g., monitoring the environment, but we also want to provide services to other parties such as citizens, for example a public hotspot, charging capabilities, and local news. Furthermore, we want to provide different services depending on whether we are in regular everyday mode or in the event of a crisis. These multiple objectives and our aim for getting the acceptance of citizens greatly distinguishes our approach and the problems we face from other monitoring or smart city solutions.
 
\section{Design}
\label{sec:design}

\begin{figure}
    \centering
    \vspace{2mm}
    \includegraphics[width=\columnwidth]{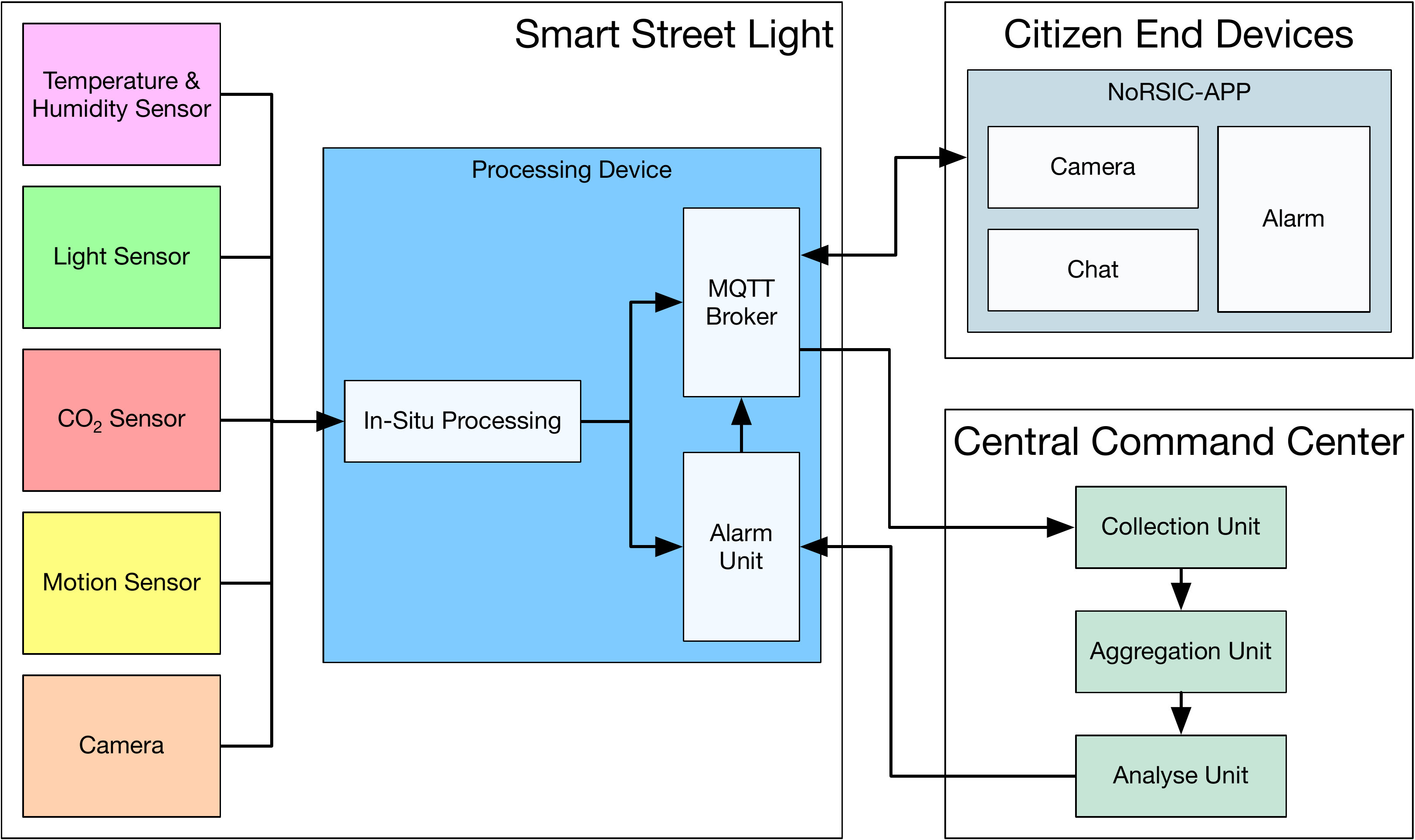}
    \caption{Architecture of our communication infrastructure including smart street light and mobile citizen app}
    \label{fig:archi}
\end{figure}

\subsection{Smart Street Light}
Smart street lights are an emerging technology for digital cities.
Manufacturers of such devices are currently exploring their possibilities and chances.
The main goal of smart street lights is to provide additional features to citizens like Internet access through integrated Wi-Fi access points (AP) or to provide additional information from local sensors.
Furthermore, using built-in technology, smart street lights can be dimmed when no activity is detected and brightened otherwise to reduce their power consumption.

Since no common feature set for a smart street light has emerged yet,
we present a set of features that we think should be an integral part of smart street lights in the future to help digital cities to cope with crises.
Our proposed architecture using smart street lights, a corresponding mobile app for citizens, and an emergency command center is shown in Fig.~\ref{fig:archi}.

First, our smart street light contains a Wi-Fi access point (AP) that provides access to Internet services, but also provides local information.
Additionally, a set of sensors should be available for various reasons.
Temperature, humidity, CO$_2$ and particulates, motion detectors, and light sensors should be available as a basic set of sensors.
Furthermore, in endangered areas with high crime rates, cameras should also be available in the lamp posts.
We also suggest including small batteries and solar panels, because lamp posts are usually set up in the open air.
In case of a crisis, a visual guidance system should be available as well, to help directing crowds remotely.
Finally, we propose having a dedicated mesh network that connects all smart street lights to a city-wide network and finally to a central data store, where the sensor data is collected, aggregated, and analyzed.

\subsubsection{Everyday Mode}
When no crisis happens or has happened (from here on called everyday mode), the smart street light functions as a light emitter and Internet AP for the inhabitants of the city.
To provide relevant local information and to increase the acceptance by citizens, the existing sensors can be used to show the temperature, humidity, and other available sensor data.
Furthermore, the CO$_2$ and particulates sensor can be used to monitor the local air quality.
Together with temperature and humidity, these four sensors are also evaluated in-situ, meaning directly at the smart street light, to give local warnings about different events.
The CO$_2$ and particulates sensor could be used to inform the local residents to close their windows due to poor air quality.
Temperature and humidity sensors are also valuable for local climate information and behavioral suggestions like giving drinking advice during hot periods.
Motion detectors and light sensors can be used to only turn on the light if it is dark enough and motion is detected, i.e., people that require light.
Additionally, the light sensor can also be used to dim or brighten the light according to the current situation.
In everyday mode, the sensor sampling rate can be rather low, since the data is only for information purposes or to detect slow changing situations like day-night changes.
Furthermore, data collected by these sensors can be used for further research with respect to the global climate change.
Finally, the sensors can also be used to detect various kinds of crises.
A high concentration of particulates and at the same time an extreme rise in temperature could, for example, indicate a fire.

The optional camera should, for privacy and data protection reasons, not record or stream data anywhere, but evaluate the taken pictures in situ.
Furthermore, computer vision and machine learning algorithms can be used to detect visual concepts locally, like fire or a crowd of people running in the same direction, indicating some sort of crisis.

Finally, we propose a central command center, where all data is gathered, stored, and analyzed.
This central command center should also have the ability to issue public warnings or alarms in case the local crisis detection failed or the other way around, to revoke a false alarm.

In everyday mode, the batteries can serve two use cases.
First, with the rise of battery-electric vehicles, solar panels charge the built-in battery, which can be used for charging a vehicle's battery using a built-in vehicle charger.
Second, decentralizing the electrical grid and placing buffer batteries to cope with temporary high energy demands is also a topic in the digital city of the future.

During everyday mode, the visual guidance system can signal different states.
When combined with charging spots for battery-electric vehicles, they can be used to indicate if the spot is available or currently out-of-order, or another vehicle is already charging.

\subsubsection{Emergency Mode}
Since our proposed smart street light includes several sensors and a camera whose data is processed locally,
it can detect crises locally and autonomously, which then triggers the emergency mode.
The functionality of a smart street light can then change to support crisis management.

During everyday mode, the Wi-Fi AP serves as the access point to the Internet.
In many crisis situations, the Internet will not be available,
but the Wi-Fi AP can still form a local network that is connected to the central command center.
Thus, a city-wide communication with citizens is potentially still possible using our smart street light approach.
In interesting option is to include smartphones of citizens to the mesh network.
Unfortunately, smartphone vendors nowadays do not allow ad-hoc mesh networks or any other kind of wireless communication apart from the regular AP mode.
Even with custom operating systems or special crisis apps, most users will not have them installed when crises happen and retrofitting all users after the event happened is not feasible.
Using smart street lights, we can handle the situation well.
The built-in sensors can now help to provide a better picture of what is going on.
First responders can plan and coordinate their missions based on the information of the sensors. For example,
a smart light may have announced a fire, but the fire department may not know where the fire is taking place.
Seeing the sensor values coming from a particular lamp post and the corresponding values can help to estimate the size, intensity, and location of the fire.
To further improve the granularity of the information, the sampling rate of the sensor may be increased, helping to understand the situation better.

Furthermore, the camera that only evaluates taken pictures locally during everyday mode, will now send the pictures to the central command center, again, supporting the emergency management staff to handle the situation.

A further type of functional morphing is performed by the mesh network.
During crises, the different rescue helpers and authorities on site need a common communication channel to coordinate recovery actions.
This can be provided by the mentioned city-wide wireless mesh network.
In addition, this communication channel must not interfere with other legitimate spectra to prevent disturbances,
and it should be secure and robust against attackers, e.g., during cyber attacks or an act of terrorism.
Therefore, we propose to morph the functionality of the wireless mesh network to address these properties.

In emergency mode, the visual guidance system can be used to direct people.
A red signal could mean that no one should go this way, and green signal, on the other hand, could indicate a safe direction.

Finally, since a smart street light is powered by solar energy and a battery, it can be powered for a long time, even if the electrical supply is damaged or not available. Furthermore, to sustain the citizens' ability to communicate, charging wireless devices from a street light is another viable option during a crisis.
Due to the wireless mesh network, every smart light can operate in a self-sustaining manner in the first phases of a crisis but still communicate with first responders and provide a secure channel for them.


\subsection{Mobile Services and Citizen App}

We complement the infrastructure created by the smart street lights with a software infrastructure, a middleware, to create citizen applications that are resilient during crises.
Our middleware for citizen applications is based on web standards, enabling a zero-installation distribution on all current end-user platforms, and automatically provides suitable resilient communication adapting to the current situation.
Maximum compatibility is an important requirement for our use case, to ensure that the application is available to every citizen in case of a crisis.


\subsubsection{Everyday Mode}

During everyday mode, the example application connects to a city-wide central server to acquire a feed of local information. 
We use city-wide news, similar to RSS or local twitter feeds, as an example. Information can further include city-wide traffic information, public transport, availability of services such as bike-sharing, and other services of the digital city.
Providing functionality even during everyday mode is necessary to give citizens a strong incentive for already using the application before a crisis, to ensure they are familiar with its usage, and to make it readily available.
We use a client-server topology for everyday mode to ensure low latency of updates, high quality of the provided information, and low communication overhead.
The access points of the smart street light function as caches for information from the central server, reducing duplicate traffic when multiple citizens are connected to the same access point.

\subsubsection{Emergency Mode}

Our middleware switches the application to emergency mode, when instructed to do so by the crisis detection mechanism of a local smart street light, or when the application itself encounters permanent connection issues attributed to an emergency situation.
Application developers do not have to explicitly support emergencies. Instead, the middleware continues to provide the same information on a best-effort basis in emergency situations, to ensure the application can continue to function normally -- possibly at reduced service quality.
To this end, a number of alternative communication schemes are established during emergency mode.
The application on the citizen device starts to exchange information directly with the access point of the smart street light.
Information is exchanged in both directions, i.e., both the citizen device obtains updates from the smart street light and the street light is updated by the citizen device if the device has newer information. The smart street light relays such information to other connected devices.

For example, in a crisis situation, only a few mobile phones may still enjoy connectivity to a cellular network to receive new information.
In such a situation, a communication path is established from the cellular network through the connected mobile phone and the smart street light to all other mobile phones connected to the street light.
Furthermore, our middleware enables individual devices to opportunistically establish direct connections with other local devices to ensure that communication is always possible.
Independently of how relevant information is acquired, the information is made available to the application in a uniform manner, abstracting over the underlying communication schemes.
The uniformity alleviates application developers from the burden of developing, testing, and maintaining different applications for everyday and emergency mode.

In addition to providing steady functionality with limited communication, application developers may provide functionality specific for emergency use cases.
For example, instead of only consuming information about the smart city, the application enables citizens to communicate with each other and with first responders in emergency situations.
Citizens may locally disseminate information that is highly relevant for their current area (e.g., the location and the nature of emergency sites, or where help is needed or available) and which can no longer be made available through a the central server due to the collapse of everyday communication infrastructure.
In our example application, first responders have special permissions to share global information, which is distributed from application to application using any of the aforementioned communication channels, based on the described technology for resilient communication.

\section{Implementation}
\label{sec:impl}

\subsection{Smart Street Light}
As a proof-of-concept implementation, we built a prototype of a smart street light and a corresponding mobile app.

\begin{figure}
    \centering
    \includegraphics[height=.92\textheight]{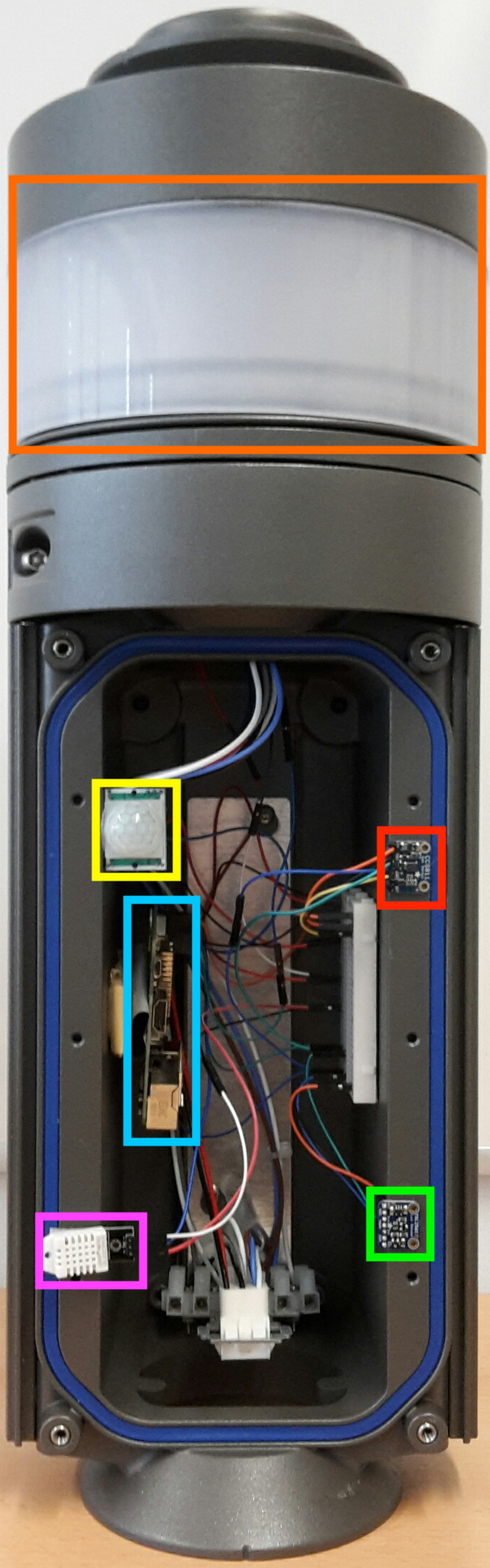}
    \caption{Sensors of the smart street light; orange: visual guidance system; yellow: motion sensor; red: CCS811 CO$_2$ sensor; blue: Raspberry Pi; green: TSL2561 light sensor; pink: AM2302 temperature and humidity sensor}
    \label{fig:lamp_prototype}
\end{figure}

We used a prototypical smart street light from Schréder\footurl{https://www.schreder.com/}, one of the largest street light vendors globally.
Their product \textit{Shuffle} is a modular street light, with some capabilities required already available.
First, their modular street light provides the street light itself and a Wi-Fi AP.
Wi-Fi is present in most commodity hardware. 
Thus, the built-in Wi-Fi AP can be used for both the Internet AP for citizens and for the mesh network between the street lights without requiring dedicated hardware. 
For special purpose solutions or other use cases, radio links such as LoRa, ZigBee, or Sigfox can easily be added to the setup.
Furthermore, Schréder also provides a camera module that we used for building the proposed prototype.
Additionally, a visual guidance system is available (orange in Fig.~\ref{fig:lamp_prototype}), containing a RGB-LED ring that can be programmed using the DALI protocol.

We additionally integrated a Raspberry Pi (blue) and a set of sensors, as shown in Fig.~\ref{fig:lamp_prototype}.
Our prototype uses an Aosong AM2302 temperature and humidity sensor\footurl{https://www.sparkfun.com/datasheets/Sensors/Temperature/DHT22.pdf} (pink), a TAOS TSL2561 light sensor\footurl{https://cdn-shop.adafruit.com/datasheets/TSL2561.pdf} (green) that can sense infra-red light and the remaining broadband spectrum.
To sense CO$_2$, an AMS CCS811 gas sensor\footurl{https://cdn.sparkfun.com/assets/learn_tutorials/1/4/3/CCS811_Datasheet-DS000459.pdf} (red) is used.
Finally, as a motion sensor, the HC-SR501 PIR sensor\footurl{https://www.mpja.com/download/31227sc.pdf} (yellow) is used.
During everyday mode, the sensors have a sampling rate of one value per 30 seconds per sensor and is increased to one sensor value per five seconds per sensor in the emergency mode.

We implemented reading, evaluating, and publishing the sensor values using Node-RED, a tool for developing and implementing dataflows visually using the flow-based programming paradigm.
It provides a level of abstraction, where no programming or computer science skills are required, thus tech savvy employees should be able to implement their own sensor evaluation software.
Furthermore, to publish data, we use the MQTT protocol that is able to distribute data with quality of service guarantees, even in unpredictable network conditions and intermittent connectivity.

To build the inter-light mesh network, we employed ESP8266 boards (ESPs from now on), using the ESP-NOW library\footurl{https://www.espressif.com/en/products/software/esp-now/overview} for providing a reliable mesh network between the street lights and the central command center.
To implement the communication channel for helping authorities, we exploited some features of the ESPs boards.
The crystal of the ESPs oscillates with a frequency of either 40 MHz or 26 MHz.
To adjust this frequency to the frequency required for emitting IEEE 802.11 compliant electro-magnetic signals, the ESPs have two Phase-Locked Loops (PLLs) (a) RFPLL for adjusting the center frequency, and (b) BBPLL for adjusting the bandwidth frequency and other peripherals.
BBPLL is adjusted using registers in the CPU, since they have to be adjusted depending on whether the ESP has a crystal with 40 MHz or 26 MHz.
This means that this factor can be set at runtime, resulting in altered bandwidths.
Since in the IC of a Wi-Fi chip different components are used to ensure a steady and clean radio signal like high- and low-pass filters, the factor for BBPLL cannot be set arbitrarily.
After analyzing the possible values, we settled for a bandwidth of 8 MHz, which is not IEEE 802.11 compliant any more.
This results in Wi-Fi signals only decodable by other ESPs with the same BBPLL values, or software-defined radios.
In fact, off-the-shelf Wi-Fi devices cannot even see the Wi-Fi frames.
Another advantage of this approach is that we are effectively shrinking the bandwidth of the Wi-Fi spectrum, reducing the overlapping space of neighboring channels and thus reducing interferences with them.
Therefore, we also favor 8 MHz over the other possible bandwidth of 16 MHz.
Since we use a layer of encryption on the MAC layer, we effectively have created a secure and hidden communication channel that reduces the interferences with legitimate radio spectra.

The built-in camera has a sample rate of one picture per second.
To detect crises locally, we trained a simple detection model to recognize an orange circle within these pictures as a proof-of-concept local detection algorithm.
If an orange circle is detected, functional morphing to the crisis mode is triggered, which increases the sensor's sampling rate to 1 sensor value per five seconds per sensor.
Additionally, a push notification is sent to all connected smartphones to inform citizens about the emergency event.

\subsection{Mobile Services and Citizen App}

Our middleware is implemented by extending the Scala programming language as a library.
We provide a general high-level declarative programming API for interactive applications using functional reactive programming~\cite{Salvaneschi2014}.
The API and implementation supports multi-threading~\cite{Drechsler:2018:TRP}, distribution~\cite{Drechsler2014}, flexible communication specifications~\cite{Weisenburger:2018:DSD}, and fault tolerance~\cite{mogk_et_al:LIPIcs:2018:9206}.
We selected Scala because it offers strong typing, flexible syntax, and a wide range of compilation targets.
Strong typing enables our API to catch many errors at compile time, which is important because applications must be resilient to crises, which in turn is a rarely tested code path.
The flexible syntax allows us to provide a unified API for user interfaces and network communication.
By integrating these concepts, the middleware is enabled to make flexible decisions on the network level, while ensuring interoperability with the displayed information.
Compiling to the web is essential to enable zero-installation distribution of the applications, and allows developers to reuse existing JavaScript libraries and techniques for UI development.
In addition, we also compile the same code to the JVM running on the central servers, ensuring interoperability and reducing required code duplication.

Normal deployment of applications is performed via HTTPS in everyday mode, users just visit a web page and can immediately start using the application.
The smart street light is able to cache local copies of the application and distribute them to local citizens.
With a typical size of less than 1\,MB, it is even often feasible to install applications over mobile data networks.
Technically, users may also directly connect devices to acquire the application manually, however the UI of current mobile devices for such a use case is very poor and outside of our control.

\begin{figure}
\vskip .3cm
\centering
\def\svgwidth{\columnwidth}
\scalebox{0.9}{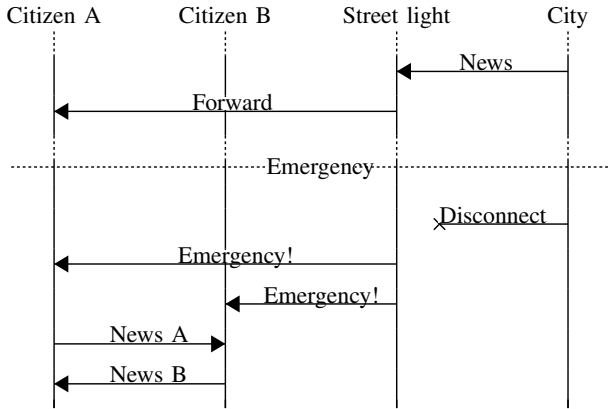}
\caption{Communication in our middleware}
\label{fig:application:communication-diagram}
\end{figure}

Figure~\ref{fig:application:communication-diagram} shows an example message flow of the middleware.
First, communication happens via a central server, and during an emergency the applications are communicating directly with each other.

Communication during everyday mode happens via WebSockets communicating with the central server.
The application additionally connects to the MQTT broker of the smart street light to acquire local information, such as the emergency signal.
When the emergency mode is activated, the middleware uses the current connections to the central server (either WebSockets or MQTT) to discover other devices and establish direct WebRTC connections for improving local communication performance.

If there is no prior connection, e.g., because the mobile device was powered on after all connections became unavailable, then devices can be manually connected to each other by scanning QR codes, or exchanging files on another ad-hoc communication medium.
The user experience for direct communication without a previously established connection on the web is the largest limitation of our approach, and our current solutions are merely workarounds until more user-friendly solutions are directly available.

With all communication mechanisms, the middleware ensures eventual consistency for the application state. 
Application developers design their application using our provided abstractions for time-changing values, and the middleware ensures that all participants eventually see the same value on their device.
Technically, eventual consistency is achieved by limiting the API for state mutation, such that all concurrent modifications can be consistently merged.
To this end, our solution incorporates state-based CRDTs~\cite{Shapiro2011}, of which efficient implementations are known~\cite{Enes2018ESC}.
In our example application, we limit potential abuse for communication by only automatically forwarding messages from known participants, such as first responders, in the network.
These known participants are determined by pre-shared public keys acquired during everyday operations.
Other messages must be reviewed by the current user of the device to be forwarded to other devices.

In general, emergency mode requires additional resources as compared to everyday mode, such as storage and network from the devices of citizens,
because those devices have to replicate some of the unavailable infrastructure.
However, we only assume emergencies to last for a short time, a couple of days at most, after which normal operations resume, and additional resources are freed again. 

\section{Experimental Evaluation}
\label{sec:eval}

In the following, we present an experimental evaluation of some of the key components of our approach. First, the covert channel, the power consumption, and the cost of the components are evaluated, and then the performance regarding message round trip times of the citizen app is presented. Cost, power consumption, and performance of the used processors, such as the ESP and Raspberry Pi, and some of the sensors, were also evaluated in previous work~\cite{baumgartner2018envmonitoring}. The feasibility  of on-device image recognition was also shown in this previous work. 

\subsection{Rescue Covert Channel}
The proposed rescue covert channel is established by downclocking the ESPs BBPLL.
Besides the already discussed advantages, downclocking the BBPLL results in a more narrow bandwidth, which could result in a lower transmission speed.
The IEEE 802.11 standard defines bandwidths of 5, 10, 20, 40 and 80 MHz, whereas the IEEE 802.11n standard only uses a bandwidth of 20 MHz on the 2.4 GHz spectrum.

\begin{figure}
    \centering
    \vspace{0.2cm}
    \includegraphics[width=\columnwidth]{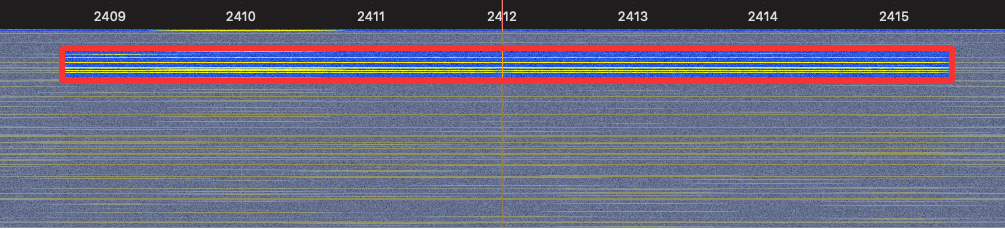}
    \caption{Waterfall view of the downclocked Wi-Fi signals}
    \label{fig:waterfall}
\end{figure}

Using our proposed approach for creating the rescue covert channel by downclocking the ESPs BBPLL results in about 8 MHz bandwidth, as shown in Fig.~\ref{fig:waterfall}, which is 40\% of the available bandwidth.

To evaluate the available and usable bandwidth measured in transmitted data per time period, we conducted an experiment where two ESPs were about 1 meter, 10 meters, and 50 meters apart. The latter is about the distance between deployed street lights, depending on the regulatory conditions and region.
We used a file of 1 MB and measured the time it took to transmit the file.
Besides the downclocked 8 MHz bandwidth, we also evaluated the standard 20 MHz bandwidth and the second option available, 16 MHz.

\begin{figure}
    \centering
    \includegraphics[width=0.9\columnwidth]{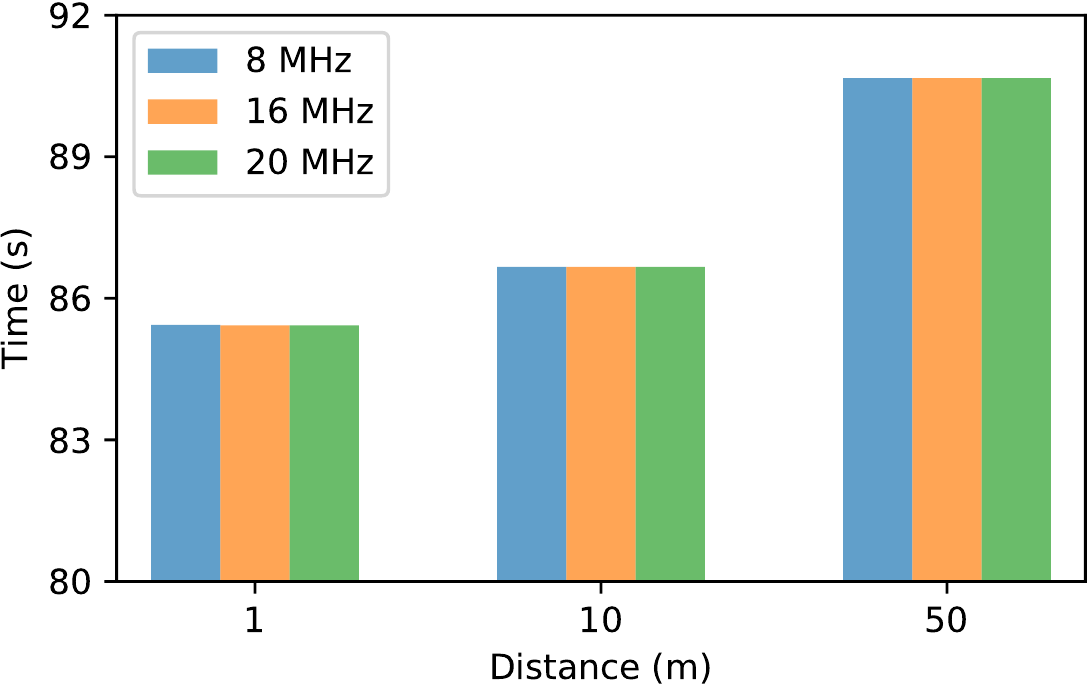}
    \caption{Transmission time in seconds it takes for sending a 1 MB file between two ESPs.}
    \label{fig:cover_eval}
\end{figure}

As shown in Fig~\ref{fig:cover_eval}, the transmission time is not related to the bandwidth but only depends on the distance between the two ESPs.
The standard error is also below 0.5 seconds for all experiments (not shown here).
Although IEEE 802.11n can achieve 150 MBit/s per used antenna pair, the ESPs are not able to reach these speeds due to their low-power architecture.
It can be seen that transmitting 1 MB over a distance of 1 m takes about 85 seconds on the average.
This results in about 96 kbit/s transmission speed.
Since this is about 0.06\% of the achievable IEEE 802.11n speed, narrowing the bandwidth does not impact the transmission speed.
For a distance of 10 meters, this results in 92 kbit/s, and for a distance of 50 meters in about 88 kbit/s.

All sensor readings consist of 32 bit floating point numbers.
Our proposed approach requires six sensors (motion, infrared, broadband light, temperature, humidity, CO$_2$), which results in 192 bit, sampled every 5 seconds in emergency mode, which is about 0.0384 kbit/s.
This leaves enough room for about 2,000 smart street lights within a single hop range to fill the available 88 kbit/s for a distance of 50 m.

If we assume that in the future hardware will be able to use 100\% of the available IEEE 802.11n bandwidth, we would still have 63 Mbp/s available even in emergency mode, which is theoretically sufficient to stream 4k videos and thus feasible for our proposed approach.

To summarize, using our proposed approach in creating the rescue covert channel, we can build a secure and hidden channel for rescuers without interfering too much with other legitimate signals with enough available bandwidth and capacity for all sensor values and city-wide communication between first responders.

\subsection{Power Consumption}

The energy requirements of the used components are rather low.
The ESP needs less than 200 mA\footnote{https://quadmeup.com/esp8266-esp-01-low-power-mode-run-it-for-months/} when sending data over Wi-Fi and below 70 mA when idling, while the used sensors sum up to another 100 mA when measuring and about 1 mA when idle.
Using a 5 V power supply, this results in additional 1.5 W, which is negligible compared to the remaining power consumption, e.g., introduced from the light source itself (usually between 20 and 100 W) or the Wi-Fi AP. More detailed numbers regarding power consumption can be found in our previous work \cite{baumgartner2018envmonitoring}.

\subsection{Cost Estimation}

The cost for a traditional street light highly depends on the country and the region where it is deployed, the used light source (LED, HPS, gas, etc.), and additional accessories like solar panels or cameras.
In the US, the prices range from about \$1,500 to \$10,000.
The sensors used in our work are available for about \$15 in total, excluding a camera which came with the smart street light. The ESP for the covert channel costs less than \$1.
For production versions of the smart street light, the Raspberry Pi would not be required, since Schredér ships its own head unit, where the logic can be implemented.
However, even when including a Raspberry Pi, the overall cost of the components is less than \$100. Adding additional cameras, external housing and a PSU with battery buffer for upgrading a traditional street light still lets the total costs stay below \$500.
Compared to the cost for the street light itself, the cost for the components required for our solution are negligible.



\subsection{Citizen App}

\begin{figure}
\centering
\vspace{3mm}
  \includegraphics[width=.8\columnwidth]{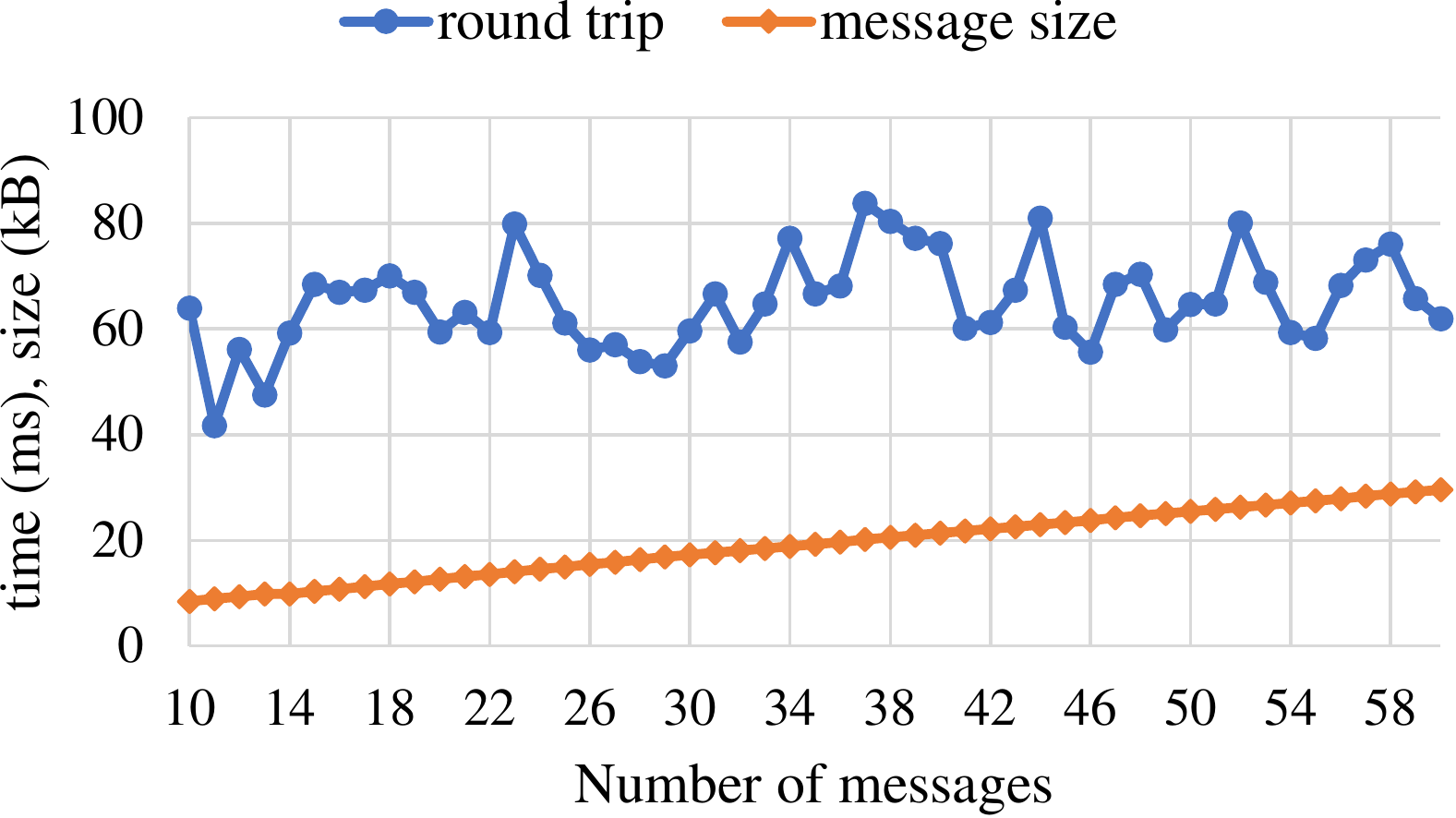}

  \caption{Round trip times and message sizes for increasingly more messages.}
  \label{fig:evaluation:application}
\end{figure}

We evaluate our example news application with respect to message sizes and message delays in Figure~\ref{fig:evaluation:application}.
We start at 10 news articles (number of messages) and continually add messages to the shared state of the application.
We measured the time it takes for a single remote mobile device to acknowledge that a new article was received, processed, displayed to the user, and forwarded to other devices in the network.
The combined time (round trip) of these operations stays under 100\,ms during our measurements, and most of that time is spent on actually processing the messages.
Due to the way synchronization of our infrastructure works and to ensure that all participants eventually have all state available, message sizes continually grow when more state is added.
This size can be reduced by purging old messages, e.g., once they become irrelevant because they are out of date.

\section{Conclusion}
\label{sec:concl}

In this paper, we presented a novel communication infrastructure to increase the resilience of digital cities in the presence of crisis events. Citizens not only benefit from our approach during a crisis, but also during everyday life. Our approach is based on smart street lights that provide automatic crisis detection, environmental sensing, as well as public services for tourists and citizens alike. During a crisis, the functionality of the smart street light can morph according to the situation and provide live video, high sampling frequency sensing, and a secure covert channel for communication with a command center. Furthermore, a novel platform-agnostic web app complements the street light functionality to provide general information during everyday life and continues to work without Internet in the event of a crisis. The app allows citizens to share messages and images with others using device-to-device communication. Our experimental evaluation demonstrated the feasibility of our approach.

In the future, we will extend the sensory capabilities of the smart street light to further enhance local situation assessment and thus improve the information about the current situation that is provided to citizens.
Furthermore, we will investigate communication mechanisms to improve communication between end devices dealing with a (partially) collapsed communication infrastructure based on approaches that can seamlessly transition within the continuum between fully centralized communication schemes (when possible) and decentralized ones (when necessary).
Finally, we plan to conduct user studies to evaluate the accessibility of our web app and gain insights into how communication technology can help citizens in mastering emergency situations and -- based on the results -- refine our app's design to improve the effectiveness of our approach to support citizens during crises.




\section*{ACKNOWLEDGMENT}

This work is funded by the LOEWE initiative in Hessen, Germany (emergenCITY and Natur 4.0), and the Deutsche
Forschungsgemeinschaft (SFB 1053 and ME 1407/16-1).

\bibliographystyle{IEEEtran}
\bibliography{IEEEabrv,literature}

\end{document}